\documentstyle[12pt]{article}
\begin{document}

\setlength{\textheight}{240mm}
\voffset=-25mm
\baselineskip=20pt plus 2pt

\begin{center}

{\large \bf On the Energy of Vaidya Space-time}\\
\vspace{5mm}
\vspace{5mm}

I-Ching Yang{\footnote{E-mail:icyang@nttu.edu.tw}} 

Systematic and Theoretical Science Research Group,\\
and Department of Natural Science Education, \\ 
National Taitung University, Taitung, Taiwna 950, Republic of China\\

\end{center}
\vspace{5mm}

\begin{center}
{\bf ABSTRACT}
\end{center}

In this paper we calculate the energy distribution of six cases of Vaidya-type solutions 
in the M{\o}ller prescription.  With the exception of the energy complex of M{\o}ller for 
the monopole solution which vanishes everywhere, the other solutions have a non-zero energy 
component. Only the energy distributions of the de Sitter and anti-de Sitter solution are 
independent of the advanced/retarded time $v$.  For the radiating dyon solution, the 
difference in the energy complex between M{\o}ller's and Einstein's prescription is just 
like the case for the Reissner-Nordstr\"{o}m solution.

\vspace{2cm}
\newpage

A non-static spherically symmetric solution of Einstein's equations for an imploding
(exploding) null dust fluid is found by Vaidya in 1951~\cite{1}.  In various contexts this
``null dust" may be interpreted as a high-frequency electromagnetic or gravitational wave,
incoherent superposition of aligned waves with random phases and polarisations, or as
massless scalar particles or neutrinos.  Since then, the solution has been intensively 
studied in gravitational collapse~\cite{2}.  In particular, Papapetrou~\cite{3} firstly 
showed that this solution can give rise to the formation of naked singularities, and thus 
provided one of the earlier counterexamples to the cosmic censorship conjecture~\cite{4}.  
Furthermore, the solution was generalized to the charged case~\cite{5}, and the charged 
Vaidya solution has been studied soon in various situations.  It had been used to study the 
thermodynamics of black holes by Sullivan and Israel~\cite{6} and be a classical model for 
the geometry of evaporating charged black holes by Kaminga~\cite{7}.  In the meantime, 
Lake and Zannias~\cite{8} studied the self-similar case and found that, similar to the 
uncharged case, naked singularities can be also formed from gravitational collapse.  
Otherwise, Chamorro and Virbhadra~\cite{9} obtained the dyonic case, in which both with 
the electric and magnetic charge, and Husian~\cite{10} further generalized the Vaidya 
solution to a null fluid with a particular equation of state.  Husian's solutions have 
been lately used as the formation of black holes with short hair~\cite{11}.

In this article, we would to consider the energy of several well-known cases of Vaidya-type
solution.  The energy in curved space-time has been a subject of extensive research since
the early days of general relativity.  Here, to evaluate the energy of a system in general 
relativity is for two physical reasons.  First, the total energy of a system must be a 
conserved quantity and could play an important role in solving the equation of motion.  
Second, the energy distribution must be positive everywhere if attractive force exists only.  
Early energy-momentum investigations for gravitating systems gave reference-frame-dependent 
energy-momentum complexes, and many physicists such as Einstein~\cite{12}, Tolman~\cite{13}, 
Landau and Lifshitz~\cite{14}, Papapetrou~\cite{15}, Bergmann and Thompson~\cite{16}, 
Weinberg~\cite{17}, and M{\o}ller~\cite{18}, had given different definitions for the 
energy-momentum complex.  The M{\o}ller energy-momentum complex is an handy tool for the 
calculation of the energy-momentum localization and allows obtaining satisfactory results 
for the energy and momentum distributions in several cases of general relativistic 
systems~\cite{19,20,21}. The M{\o}ller energy-momentum complex in a four-dimensional 
background is given as~\cite{18}
\begin{equation}
\Theta^{\mu}_{\nu} = \frac{1}{8\pi} \frac{\partial \chi^{\mu\sigma}_{\nu}}
{\partial x^{\sigma}} ,
\end{equation}
where 
\begin{equation}
\chi^{\mu\sigma}_{\nu} = \sqrt{-g} \left( \frac{\partial g_{\nu\alpha}}
{\partial x^{\beta}} - \frac{\partial g_{\nu\beta}}{\partial x^{\alpha}}
\right) g^{\mu\beta} g^{\sigma\alpha} 
\end{equation}
is the M{\o}ller's superpotential, a quantity antisymmetric in the indices $\mu$, 
$\sigma$.  According to the definition of the M{\o}ller energy-momentum complex, the energy 
within a volume V is given as 
\begin{equation}
E= \int_V \Theta^{0}_{0} d^3 x =\frac{1}{8\pi} \int_V \frac{\partial 
\chi^{0k}_{0}}{\partial x^{k}} d^3 x,
\end{equation}
and the Latin index takes values from 1 to 3.

The line element, in terms of the standard spherical coordinates, of the non-static 
spherically symmetric type D solution~\cite{22} can be expressed as 
\begin{equation} 
ds^2 = e^{2\psi(v,r)} \left[ 1-\frac{2m(v,r)}{r} \right] dv^2 - 2\epsilon e^{\psi(v,r)} dvdr - r^2 d\theta^2 - r^2 \sin^2 \theta d\phi^2 , 
\end{equation}
note that $m(v,r)$ is usually called the mass function and relates to the gravitational 
energy within a given radius $r$.  When $\epsilon=+1$, the null coordinate $v$ represents 
the Eddington advanced time, in which $r$ is decreasing towards the future along a ray 
$v=constant$ (ingoing).  On the other hand, when $\epsilon=-1$, it represents the Eddington 
retarded time, in which $r$ is increasing towards the future along a ray $v=constant$ (outgoing).  
In the following, we shall consider the particular case where $\psi(v,r)=0$, and the stress-energy 
tensor of radiation and null dust fluid in Vaidya solution is of the form~\cite{22}
\begin{equation}
T_{\mu \nu}=(\rho +P)(l_{\mu} n_{\nu} + l_{\nu} n_{\mu}) +P g_{\mu \nu} +\mu l_{\mu} l_{\nu} ,
\end{equation}
where
\[ \mu = \frac{\epsilon \dot{m}(v,r)}{4\pi r^{2}},\;\; \rho = \frac{m'(v,r)}{4\pi r^{2}},\;\;
P = - \frac{m''(v, r)}{8\pi r} \]
and 
\[ \dot{m}(v,r) \equiv \frac{\partial m(v,r)}{\partial v},\;\; m'(v,r) \equiv \frac{\partial m(v,r)}{\partial r}. \]
Here $l_{\mu}$, $n_{\mu}$ are two null vector,
\[ l_{\mu}=\delta^{0}_{\mu},\;\; 
n_{\mu} = \frac{1}{2} \left( 1- \frac{2m(u,r)}{r} \right) \delta^{0}_{\mu} - \delta^{1}_{\mu}, \] 
\[ l_{\lambda} l^{\lambda} =  n_{\lambda} n^{\lambda} =0,\;\; l_{\lambda} n^{\lambda} =-1 .\]
Therefore, the nonvanishing component of M{\o}ller's superpotential is 
\begin{equation}
\chi^{01}_0 = \frac{\sin \theta}{\epsilon} (2m -2m'r) .
\end{equation}
Applying the Gauss theorem to (3) and using (6), we evaluate the integral over the surface of 
a sphere with radius $r$, and find that the energy distribution is as the form
\begin{equation}
E= \frac{1}{\epsilon} (m - m'r) .
\end{equation}

Next, The following cases include six known Vaidya-type solutions of the Einstein 
field equations with spherical symmetry: \\ 
(i) The monopole solution~\cite{23}: The mass function of the monopole solution is given as
\begin{equation}
m(v,r) =\frac{ar}{2}  ,
\end{equation}
where $a$ is an arbitrary constant.  The corresponding solution can be identified as representing
the gravitational field of a monoploe.  After the previous computatuion, all components of 
M{\o}ller's superpotential is zero, and the energy component of M{\o}ller energy-momentum complex 
is obtained with
\begin{equation}
E=0  .
\end{equation}
(ii) The de Sitter and Anti-de Sitter solutions~\cite{22}: The well-known de Sitter and Anti-de Sitter
solutions in the Vaidya's radiation coordinates show that the mass function is
\begin{equation}
m(v,r) =\frac{\Lambda}{6} r^3 ,
\end{equation}
where $\Lambda$ is the cosmological constant.  This corresponds to the de Sitter solution for
$\Lambda >0$ and to anti-de Sitter solution for $\Lambda <0$.  The required nonvanishing
component of $\chi^{0k}_{0}$ is 
\begin{equation}
\chi^{01}_{0} = -\frac{\sin \theta}{\epsilon} \frac{2}{3} \Lambda r^3   ,
\end{equation}
and the obtained energy component within a sphere of radius $r$ is  
\begin{equation}
E= - \frac{\Lambda}{3\epsilon} r^3  .
\end{equation}
(iii) The charged Vaidya solution~\cite{22}: For the charged Vaidya solution, the mass function is found as
\begin{equation}
m(v,r) =f(v) -\frac{q^2(v)}{2r} ,
\end{equation}
where the two arbitrary functions $f(v)$ and $q(v)$ represent, respectively, the mass and electric 
charge at the advanced (retarded) time $v$.  The non-zero component of M{\o}ller's superpotential
only is obtained as
\begin{equation}
\chi^{01}_{0} = \frac{\sin \theta}{\epsilon} \left[ 2f(v) -\frac{2 q^2 (v)}{r} \right] ,
\end{equation}
and the energy component within a shpere of radius $r$ of M{\o}ller energy-momentum 
complex is evaluated with
\begin{equation}
E = \frac{1}{\epsilon} \left[ f(v) -\frac{q^2 (v)}{r} \right] .
\end{equation} 
(iv) The monopole-de Sitter-charged Vaidya solutions~\cite{22}: According to the monopole-de Sitter-charged 
Vaidya solutions, its mass function is represented as  
\begin{equation}
m(v,r) =\frac{ar}{2} +\frac{\Lambda}{6} r^3 +f(v) -\frac{q^2(v)}{2r}  .
\end{equation}
Using the definition of M{\o}ller energy-momentum complex, the nonvanishing component of M{\o}ller's
superpotential is
\begin{equation}
\chi^{01}_{0} = \frac{\sin \theta}{\epsilon} \left[ -\frac{2}{3} \Lambda r^3 +2f(v) -\frac{2 q^2 (v)}{r} \right]  ,
\end{equation}
and the energy component within a shpere of radius $r$ of M{\o}ller energy-momentum 
complex is
\begin{equation}
E = \frac{1}{\epsilon} \left[ -\frac{\Lambda}{3} r^3 +f(v) -\frac{q^2 (v)}{r} \right] .
\end{equation}
(v) The radiating dyon solution~\cite{9}: The metric, which describes the gravitational field of non-rotating 
massive radiating dyon, is found by Chamorro and Virbhadra, and its mass function is
\begin{equation}
m(v,r) =M(v) - \frac{q_e^2 (v) +q_m^2 (v)}{r} .
\end{equation}
With the energy-momentum pseudotensor of M{\o}ller, we obtain the nonvanishing component
of M{\o}ller's superpotential is
\begin{equation}
\chi^{01}_{0} = \frac{\sin \theta}{\epsilon} \left[ 2M(v) -\frac{2 q_e^2 (v) +2 q_m^2 (v)}{r} \right] ,
\end{equation}
and the energy within a shpere of radius $r$ is calculated as
\begin{equation}
E= \frac{1}{\epsilon} \left[ M(v) -\frac{q_e^2 (v) + q_m^2 (v)}{r} \right] .
\end{equation}
(vi) The Husain solutions~\cite{10}: This solution found by Husian with imposing the equation of state $P=k\rho$ have
the mass function defined as
\begin{equation}
m(v,r) =f(v) -\frac{g(v)}{(2k-1)r^{2k-1}} ,
\end{equation}
where $f(v)$ and $g(v)$ are two arbitrary functions, and $k$ is a constant.  Then, by the definition of
M{\o}ller energy-momentum complex, the required non-zero component of M{\o}ller's superpotential is
\begin{equation}
\chi^{01}_{0} = \frac{\sin \theta}{\epsilon} \left[ 2f(v) +\frac{2k-2}{2k-1} \frac{2g(v)}{r^{2k-1}} \right] ,
\end{equation}
and the energy component within a shpere of radius $r$ of M{\o}ller energy-momentum 
complex is 
\begin{equation}
E= \frac{1}{\epsilon} \left[ f(v) +\frac{2k-2}{2k-1} \frac{g(v)}{r^{2k-1}} \right] .
\end{equation}

In Ref[24] it was found that energy-momentum complexes provide the same acceptable 
energy-momentum distribution for some systems. However, for other systems, these 
prescriptions are disagreed.  Based on some analysis of the results known with many 
prescriptions for energy distribution (including some well-known quasi-local mass 
definitions) in a given space-time, Virbhadra~\cite{24} remarked that the formulation by 
Einstein is still the best one, however there is really no consensus as to which is the best.  
But, in his previous paper, which motivated us to study this further, Lessner~\cite{25} argued 
that the M{\o}ller energy-momentum expression is a powerful concept of energy and momentum 
in general relativity.  However, Hamiltonian's principle helps to solve this enigma~\cite{26}.  
Each expression has a geometrically and physically clear significance associated with the 
boundary conditions.  

In this work, we briefly present the energy component of M{\o}ller energy-momentum complex for 
six Vaidya-like solutions.  Except the energy complex of M{\o}ller for the monopole solution 
vanishes everywhere, for the de Sitter, anti-de Sitter, charged Vaidya, monopole-de Sitter-charged 
Vaidya, radiating dyon and Husain solution have non-zero energy component.  Here only the energy 
distributions of the de Sitter and anti-de Sitter solution are independent on $v$.  On the case of 
radiating dyon solution, the differnece in energy complex between M{\o}ller's and Einstein's 
prescription is like the case of Reissner-Nordstr\"{o}m solution~\cite{9,21}.  For the 
Reissner-Nordstr\"{o}m space-time $E_{\rm Ein} = M - e^2 /2r$ (the seminal Penrose quasi-local mass 
definition also yields the same result agreeing with linear theory) whereas $E_{\rm M{\o}l} = M-e^2 /r$.  
Similarly, the energy of radiating dyon solution is $E_{\rm M{\o}l} = M(v) -[q_e^2 (v) +q_m^2 (v)]/r$ 
whereas $E_{\rm Ein} = M(v) -[q_e^2 (v) +q_m^2 (v)]/2r$~\cite{9}.  There is a difference of a factor 2 
in the second term.

\begin{center}
{\bf Acknowledgments}
\end{center}
This research was supported by the National Science Council of the Republic of China under contract 
number NSC 93-2112-M-143-001 and NSC 94-2112-M-143-001.

\end{document}